\documentclass[9pt,twocolumn,twoside]{article}
\usepackage{algpseudocode,graphicx}
\RequirePackage[utf8]{inputenc}
\RequirePackage[english]{babel}
\RequirePackage{amsmath,amsfonts,amssymb}
\RequirePackage{mathpazo} 
\RequirePackage[scaled]{helvet}
\RequirePackage[T1]{fontenc}
\RequirePackage{url}
\RequirePackage[colorlinks=true, allcolors=blue]{hyperref}
\RequirePackage{graphicx,xcolor}
\RequirePackage{wrapfig} 
\RequirePackage{colortbl}
\RequirePackage{booktabs}
\RequirePackage{algorithm}

\RequirePackage[left=48pt,%
right=42pt,%
top=46pt,%
bottom=60pt,%
headheight=15pt,%
headsep=10pt,%
letterpaper,twoside]{geometry}%
\RequirePackage[labelfont={bf,sf},%
labelsep=period,%
figurename=Fig.,%
singlelinecheck=off,%
justification=RaggedRight]{caption}
\date{\small{\href{https://www.osapublishing.org/ol/abstract.cfm?uri=ol-44-5-1241}{Optics Letters, Vol. 44(5), pp. 1241-1244 (2019), 'https://www.osapublishing.org/ol/abstract.cfm?uri=ol-44-5-1241}}\\\textcopyright 2019 Optical Society of America. One print or electronic copy may be made for personal use only. Systematic reproduction and distribution, duplication of any material in this paper for a fee or for commercial purposes, or modifications of the content of this paper are prohibited.}

\title{Single-pixel imaging with sampling distributed over simplex vertices}

\author{Krzysztof M. Czajkowski, Anna Pastuszczak, and Rafa\l{} Koty\'{n}ski\\University of Warsaw, Faculty of Physics, Warsaw, 02-093, Poland}

\usepackage{xcolor} 

\begin{document}
	\maketitle
\maketitle

\textbf{We propose a method of reduction of experimental noise in single-pixel imaging by expressing the subsets of sampling patterns as linear combinations of vertices of a multidimensional regular simplex. 
This method may be also directly extended to complementary sampling.
The modified measurement matrix contains non-negative elements with patterns that may be directly displayed on intensity spatial light modulators.
The measurement becomes theoretically independent of the ambient illumination, and in practice becomes more robust to the varying conditions of the experiment.  
 We show how the optimal dimension of the simplex depends on the level of measurement noise. We present experimental results of single-pixel imaging using binarized sampling and a real-time reconstruction with the Fourier domain regularized inversion method.}

Indirect image measurement techniques called single-pixel imaging and computational ghost imaging~\cite{Baraniuk2008, PRA_78_061802_Shapiro}
contribute to many novel ideas in optics. Prospect applications of these measurement methods include spectral imaging
~\cite{Multispectral_scirep2017Jin,Multispectral_life_nphot2017Pian}, polarimetric imaging~\cite{OL_37_824_Duran,
Polarimetric_AO2018Fade}, 3D imaging~\cite{Science_Sun2013,Sun:natcommun_7_12010}, 
around-the-corner imaging, imaging through scattering media~\cite{Duran:15}, 
 spectroscopy~\cite{Spectroscopy_AO2016Starling}, pattern recognition~\cite{pastor:oc2017}, and information security~\cite{Zhang:OE-26-14578}. 

In practical experimental conditions it is necessary to pay attention to the 
low signal-to-noise ratio  of the measurement, to normalize the detection signal from a bucket detector using some kind of a reference signal,  as well as to design the sampling patterns in an optimal way. One should take into account the binary modulation of the digital micromirror device (DMD) and the image formation model~\cite{Sun:OLE-100-18,Sun:OE-24-10476}. In computational ghost imaging with structured illumination the reference signal may come from another detector that measures light intensity of the sampling functions (deterministic or pseudo-random patterns) used for illuminating the object~\cite{Ferri:PRL-94-183602,OE_20_16892_Sun}. In single-pixel imaging with a modulated structured aperture it is more difficult to control the illumination conditions of the object. Then it is still possible to introduce a normalization of the detection signal using complementary sampling~\cite{Yu:SciRep-4-5834}. This  technique resembles balanced photodetection, with two photodiodes that measure the signal reflected from a DMD in two directions by mirrors set to complementary binary patterns. A similar sequential measurement is also possible but requires doubling the number of displayed patterns~\cite{OE_21_23068_Welsh}. 
There are also other ways of improving the experimental conditions at the expense of the increased number of sampling functions. For instance in~\cite{Zhang2017a,Zhang:natcom_6_6225} a differential technique obtained from the projections of the phase-shifted complex Fourier basis was introduced, with every complex function coded using $3$ or $4$ non-negative real functions. This approach with $3$ functions may be seen as a special case of the method proposed in our paper, where we would use coding on a simplex of the order $2$. Simplex coding of complex functions or ensembles of orthogonal functions with intensity patterns have been recently proposed by Tommassi~et~al.~\cite{Tommasi:OE-20-23186}. They have shown that by using intensity-encoded structured illumination it is possible to reduce the experimental mean-square-error of the noise, suggesting also that this method
allows the study of new propagating physical quantities with the classical coherent or incoherent light field playing  the role of hidden variable~\cite{Tommasi:OE-20-23186}.
\begin{figure*}[htbp]
	\centering
	\fbox{\includegraphics[width=\linewidth]{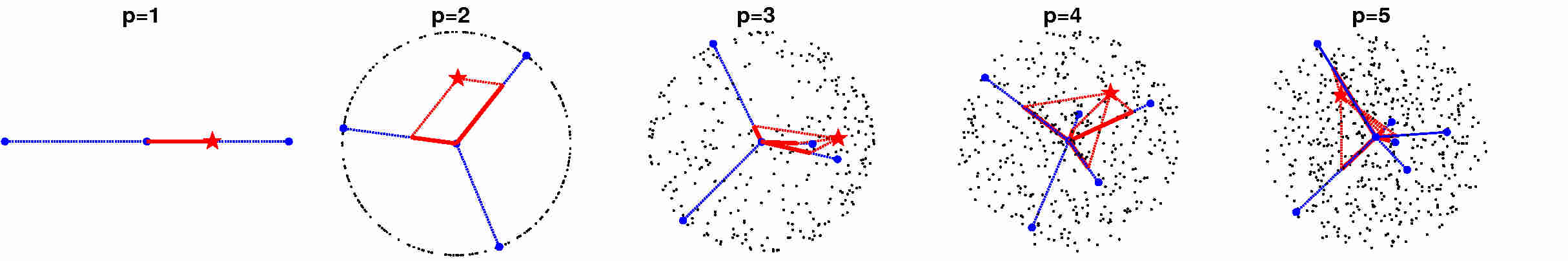}}
	\caption{Explanatory image showing how a point in a p-dimensional space marked with a star (red online) 
    may be always represented using $p+1$ non-negative coordinates only, with at least one of these coordinates equal to zero. The plot shows the projection of the $p$-dimensional space for $p=1,..,5$    onto a plane. It includes $p+1$ vertices of a regular p-simplex  (dots, blue online) which define the coordinate system, a $p$-sphere (randomly uniformly distributed small black points), and the nonorthogonal projection of point  onto $p$ out of $p+1$ axes defined with the simplex vortices nearest to that point.  
    \label{fig:nsimplex}}
\end{figure*}

In this letter we show that simplex coding reduces the influence of the slowly-varying random bias on the measurement. We also show that the optimal order of simplex coding depends on the level of noise. 
 We  show how to modify linear image reconstruction methods based on pseudoinverse or generalized inverse of the measurement matrix for use with simplex encoding and indicate that the modified method remains linear. We demonstrate that simplex encoding may be combined with complementary sampling. Finally we note that this kind of normalization method may be also used with nonorthogonal sampling. 
There is also some analogy with simplex codes used to generate pulse sequences from Hadamard functions for optical time-domain reflectometry~\cite{Jones:PTL-15-822,Lee:JLT-24-322,Wang:OE-25-5550}.

We will now briefly explain how we modify the measurement matrix by distributing its elements on a simplex. 
Suppose we have designed a real-valued measurement matrix $M$ for single-pixel imaging (complex sampling functions may be always represented by separating the real and imaginary parts).  $M$ is a $k\times n$  matrix with $k$ sampling functions stored in its rows. The number of columns $n$ equals the total number of pixels in every sampling function. We will use the vertices of a regular $p$-simplex to find a new  measurement matrix $M'$ with nonnegative elements only. The concept of the method is illustrated in Fig.~\ref{fig:nsimplex}.  For simplicity we assume that $k$ is divisible by $p$ and $k=p\cdot l$. Then the size of $M'$ will be $[k(p+1)/p, n]$.
\begin{algorithm}
	\caption{for finding the vertices of a regular simplex}\label{alg:vort}
	\begin{algorithmic}[1]
		\Function{NSimplex}{$p$}\Comment{p- the dimensionality of space}
          \State $V\gets zeros(p,p+1)$ \Comment{zero-filled $p \times (p+1)$ matrix}
          \For{$i=1\text{ to } p$}
            \State $V_{1:i-1, 1:l}\gets -(V_{1:i-1, 1:l})/i$\Comment{Rotate the vertices}
            \State $V_{i,1:i}\gets \sqrt{1-i^{-2}}$ \Comment{by $acos(-1/i)$ and}
            \State $V_{i,i+1}\gets 1$\Comment{add another axis}
          \EndFor
          \State \textbf{return} $V$\Comment{Columns of V contain vortex coordinates}
		\EndFunction
	\end{algorithmic}
\end{algorithm}
Let $V$ be a $p \times (p+1)$ matrix with the vertex coordinates $v_i$, where $i=1,..,(p+1)$, of the $p$-simplex in its columns. $V$ may be determined with Algorithm~\ref{alg:vort}, in which we iteratively add the $i$-th new orthogonal unit vector and rotate the previously determined elements by $acos(-1/i)$, where $i$ takes natural values up to $p$.  At this point we enter the procedure of finding the modified matrix $M'$ from $M$ summarized in Algorithm~\ref{alg:encode}:
For every column of matrix $M$, each set of $p$ rows defines a point $v$ 
in the p-dimensional space. Then this point is transformed into $p$ rows belonging to a single column of $M'$ by left-multiplying $v$ by the inverse of $V$ with one column removed. We choose the column to be removed in such a way  that it corresponds to the furthermost vertex with respect to $v$. The decomposition obtained this way contains only  $p$ nonnegative elements (See also Fig.~\ref{fig:nsimplex}). Therefore, for every point $v$ we select a nonorthogonal $p$-element basis using $p$ out of $p+1$ vertices of the p-simplex.
\begin{algorithm}
	\caption{for finding the modified measurement matrix with non-negative values distributed over the simplex vertices}\label{alg:encode}
	\begin{algorithmic}[2]
		\Function{SimplexSampling}{$M,V, k, p , n$}\Comment{M- real-valued k x n measurement matrix; V- vertices of a p-simplex}
          \State $l\gets k/p$ 
          \State $M'\gets zeros(l (p+1),n)$ \Comment{zero-filled $l(p+1)\times n$ matrix}
         \For{$m=0\text{ to } l-1$} \Comment{Loop over bundles of rows of M}
            \For{$c=1\text{ to } n$} \Comment{Loop over columns of M}
              \State $v\gets M_{m p+1:(m+1) p,c}$ \Comment{Extract a p-vector from  M}
              \If {$\|v\|>0$}	
                \State $q\gets argmax_{q\in 1,..,(p+1)} \| v-V_{1:p,q}\|^2$ \Comment{Find the \\ \hspace{\fill} furthermost simplex vertex with respect to v}
                \State $v\gets [V_{1:p,1:q-1},V_{1:p,q+1:p+1}]^{-1}\cdot v$ 
                \Comment{Decom- \\ \hspace{\fill} pose v in the local basis and store in M' }
                \State $M'_{m(p+1)+1:m(p+1)+q-1,c}\gets v_{1:q-1}$ 
                \State $M'_{m(p+1)+q+1:(m+1)(p+1),c}\gets v_{q:p}$ 
              \EndIf
            \EndFor
          \EndFor
          \State \textbf{return} $M'$\Comment{Return the modified measurement matrix}
		\EndFunction
	\end{algorithmic}
\end{algorithm}
Decomposition of different points $v$ extracted from matrix $M$ uses different vertices of the p-simplex, and overall we need $p+1$ rows in $M'$ to represent every $p$ rows of $M$. Storing the new representations of $v$ in $M'$ concludes the algorithm.
 
We will now confirm that in a noise free scenario the compressive measurement of an image $x$ (with pixels having nonnegative values and stored for convenience in one vector) with either $M$ or $M'$ measurement matrix are equivalent,
\begin{equation}
y=M\cdot x, \text{ or }y'=M'\cdot x.\label{eq.measurement}
\end{equation} 
The original measurement matrix $M$ may be expressed with the modified measurement matrix $M'$ as
\begin{equation}
M=Q\cdot M', \text{ with } Q=\mathbb{I}_l \otimes V,\label{eq.invrelation}
\end{equation}
where $\mathbb{I}_l$ is the identity matrix of dimension $l$ and $\otimes$ denotes the Kronecker product. We note that $Q$ is not a square matrix and is therefore not invertible. It is straightforward to calculate $M$ from $M'$ with Eq.~(\ref{eq.invrelation})  but we had to use Algorithm~\ref{alg:encode} to find $M'$ from $M$. We may add that Algorithm~\ref{alg:encode} as well as Eq.~(\ref{eq.invrelation}) remain valid for both orthogonal and non-orthogonal sampling functions.
Equations~(\ref{eq.measurement}) and~(\ref{eq.invrelation}) allow us to express $y$ with $y'$ as 
\begin{equation}
y=Q\cdot y'.\label{eq.equiv_y}
\end{equation}
Hence, for an arbitrary real-valued sampling matrix $M$ we may use equivalently matrix $M'$ with non-negative samples at the expense of increasing the number of samples by a factor of  $(1+p^{-1})$. The overhead in terms of the number of samples is between doubling this number for $p=1$ and a negligible increase by a single sample for $p=k$. 

The center of mass of a regular simplex with verticies $v_i$ is located at the center of the coordinate system $\sum_{i=1,..,(p+1)} v_i=V\cdot \mathbb{1}_{(p+1)\times 1}=0\cdot \mathbb{1}_{p\times 1}$, where $\mathbb{1}_{n\times m}$ is a ${n\times m}$ matrix filled with ones. Consequently, when Eq.~(\ref{eq.equiv_y}) is used to find $y$, any constant bias becomes removed from the measurement $y'$ in a similar way as if one performed a differential measurement,  
 \begin{equation}
  Q\cdot (y'+const)=y+const\cdot (\mathbb{I}_l \otimes V)\cdot \mathbb{1}_{l(1+p)\times 1}=y.\label{eq.dcreduction}
 \end{equation}
According to Eq.~(\ref{eq.dcreduction}), the proposed sampling eliminates any constant additive bias, for instance due to  ambient illumination, from the measurement $y'$. 
It is however important that the bias remains unchanged during the entire measurement or at least during the measurements of every $p+1$-element bundle of sampling functions. Thus, depending on how fast the ambient conditions change, one may need to put an upper limit on the value of $p$ accordingly.

\begin{figure}
\centering
\includegraphics[width=0.8\linewidth]{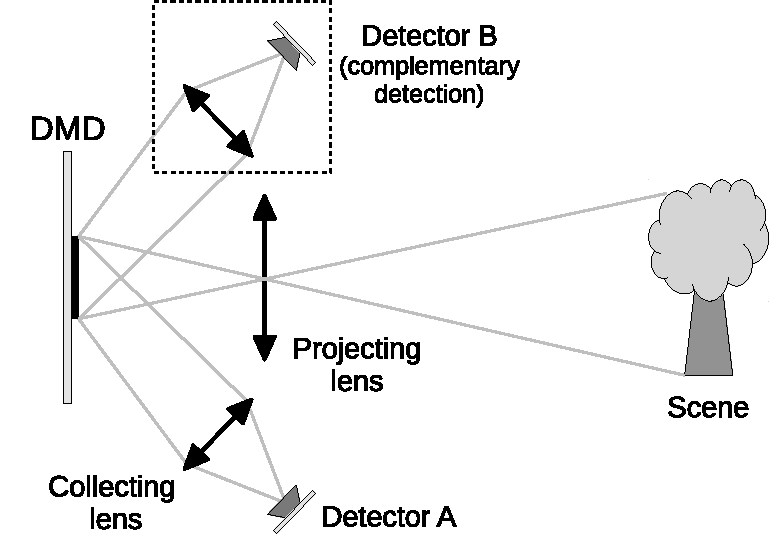}
\caption{Schematic of a single-pixel camera with optional complementary detection module (dashed line box).    \label{fig:experiment_scheme}}
\end{figure}

Commonly used sampling functions such as subsets of discrete-cosine transform (DCT), Fourier, Walsh-Hadamard, noiselet or wavelet bases have both positive and negative values (with complex-valued functions expressed as two real-valued samples), whereas light modulators such as DMD modulate non-negative intensity-based optical signals. In practical experimental conditions the measurement matrix $M$ is arbitrarily scaled and then mapped to $M_{DMD}\in[0,1]$ for displaying it on a DMD. In a straightforward approach $M=M_{DMD}-0.5$. In the presence of additive i.i.d. Gaussian detector noise $N\sim \mathcal{N}(\mu,\,\sigma^{2})$ the measurement equals
 \begin{equation}
 y^{exp}=   S \cdot \left(\begin{bmatrix}  M_{DMD}\\ \mathbb{1}_{1\times n}\end{bmatrix}\cdot x+N\right) \text{ with } S=\begin{bmatrix} 
  \mathbb{I}_n ,\frac{1}{2}\cdot \mathbb{1}_{n\times 1}
 \end{bmatrix}\label{eq.ns1}.
 \end{equation}
  Here, $\mathbb{1}_{1xn}$ is an additional measurement pattern with all DMD mirrors in the \textit{on} position. 
    An affine transform $S\cdot N$ of a multivariate normal distribution $N\sim \mathcal{N}^{k+1}(\mu,\,\sigma^{2})$ has a normal distribution of $S\cdot N\sim \mathcal{N}(S\cdot \mu,\sigma^{2} S\cdot S^{T})$. Therefore, the measurement (\ref{eq.ns1}) is affected by a correlated noise with the mean value of $\mu$ and the covariance matrix $\sigma^{2}(\mathbb{I}_n+0.25\cdot \mathbb{1}_{n\times n})$. On the other hand, in the proposed approach with simplex-coding we may assume that the matrix $M'_{DMD}$ displayed on the DMD does not require additional shifting, as the distribution over simplex allows to have $M'\in[0,1]$. Now, the measurement equals
  \begin{equation}
 y_{simplex}^{exp}=   Q \cdot \left(M'_{DMD}\cdot x+N\right)\label{eq.ns2}.
 \end{equation}
This time the measurement (\ref{eq.ns2}) is affected by the noise $\sim \mathcal{N}(Q\cdot \mu,\,\sigma^{2} Q\cdot Q^{T})$, which owing to Eq.~({\ref{eq.dcreduction}}) is a zero-mean white noise 
$\sim \mathcal{N}^{k}(0,\sigma^2 (1+p^{-1}))$. 
In most situations, a zero-mean uncorrelated noise obtained with simplex encoding is preferable as compared with the correlated noise with a non-zero mean obtained after a straightforward mapping of $M$ to positive values. Especially for non-stationary scenes, when the mean value of the noise may vary during the measurement and may be difficult to compensate using complementary detection.

 \begin{figure}
\includegraphics[width=\linewidth]{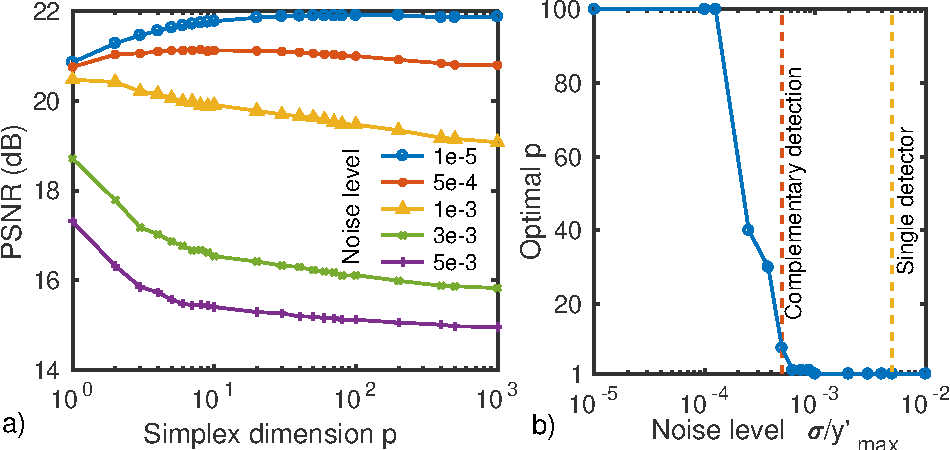}
\caption{The plot shows how to choose the dimension of the simplex for various levels of detector noise.
a) Image quality (peak signal-to-noise-ratio $PSNR(x,\tilde x)=10 log(n\cdot max(x)^2/\|x-\tilde x\|^2)$, where $x$ and $\tilde x$ are the reference and reconstructed images) vs. simplex dimension $p$ for different levels of detector noise; b) Optimal simplex dimension $p$ vs. detector noise intensity. Vertical dashed lines indicate our estimated experimental noise levels $\sigma/y_{max}\approx 5\cdot 10^{-3}$ and $5\cdot 10^{-4}$ for the single detector and complementary measurement respectively. PSNR is averaged over results for $49$ images from the USC-SIPI database. Number of sampling patterns is fixed $K=k\cdot(1+p^{-1})=2000$. The resolution is $256\times 256$. Binarized DCT sampling is used. 
\label{fig:numerical}}
\end{figure}

Image reconstruction, which is an inherent part of any single-pixel imaging measurement, can be continued with $y$ obtained from Eq.~(\ref{eq.dcreduction}) in the same way as without distributing the measurement matrix over a p-simplex. However, when the reconstruction is obtained with the Moore-Penrose pseudoinverse ($M^{\dagger}$) or generalized inverse ($M^{g}$) of the measurement matrix ( $P=M^{g}$, and $\tilde x= P\cdot y$)
one further simplification is possible
\begin{equation}
\tilde x= P\cdot y=P'\cdot y', \text{ where } P'= P\cdot Q=(Q\cdot M'_{DMD})^g \cdot Q.\label{eq.geninv}
\end{equation}
$P'$ may be precalculated before the measurement takes place, and then the reconstruction retains the linear numerical cost as a function of $n$ and $k$. 
The robustness of proposed normalization may be further improved with complementary detection 
to eliminate high frequency noise. In the experimental part of this work we use a DMD to modulate the measured images at $22$~kHz and we measure the signal with two photodiodes which at the same time capture light reflected by the DMD into two directions corresponding to the two possible states of the mirrors. A schematic of the set-up is shown in Fig.~\ref{fig:experiment_scheme}. The two signals $y_A'=M'\cdot x$ and $y_B'=(c \cdot \mathbb{1}_{k\times n}-M')\cdot x$ are subtracted and 
\begin{equation}
\tilde x=P'\cdot(y_A'-y_B'),
\end{equation}
where the constant term $\propto c$ is eliminated in the same way as in Eq.~(\ref{eq.dcreduction}).
 In  this letter we reconstruct the images after a compressive measurement with the very fast Fourier domain regularized inversion method (FDRI)~\cite{Czajkowski:OE-26-20009}. 
 The denotations for $M$, $x$, $y$, $P$ are consistent with those in~\cite{Czajkowski:OE-26-20009}. 
The scene is sampled with DCT. 
 Error diffusion is used for binarizing sampling functions, whereas simple thresholding has been used by us in~\cite{Czajkowski:OE-26-20009}. 

 In practical experimental conditions, the total number of sampling patterns 
$K=k\cdot(1+p^{-1})$ puts a limit on the operation frequency of single pixel imaging. We assume that $K=2000$, which corresponds to the frequency of $11Hz$ for the DMD modulation frequency of $22kHz$. When $K$ is fixed, the larger is the simplex dimension $p$, the larger is the dimension $k$ of the measurement $y$ and the larger is the compression ratio $k/n$. On the other hand, the smaller is $p$, the more robust is the method with respect to noise and to varying illumination conditions. Under these conditions $p$ may be optimized, and the optimal value of $p$ in terms of the reconstruction quality measured with PSNR depends on the level of additive noise. The numerical optimization results shown in Fig.~\ref{fig:numerical} indicate that for the estimated level of noise that we have in our optical set-up and for a measurement of a stationary object, we should choose $p=1$ when using a single detector, while for the complementary sampling it is the best to take a larger value of $p$.

\begin{figure}
\centering
\includegraphics[width=0.65\linewidth]{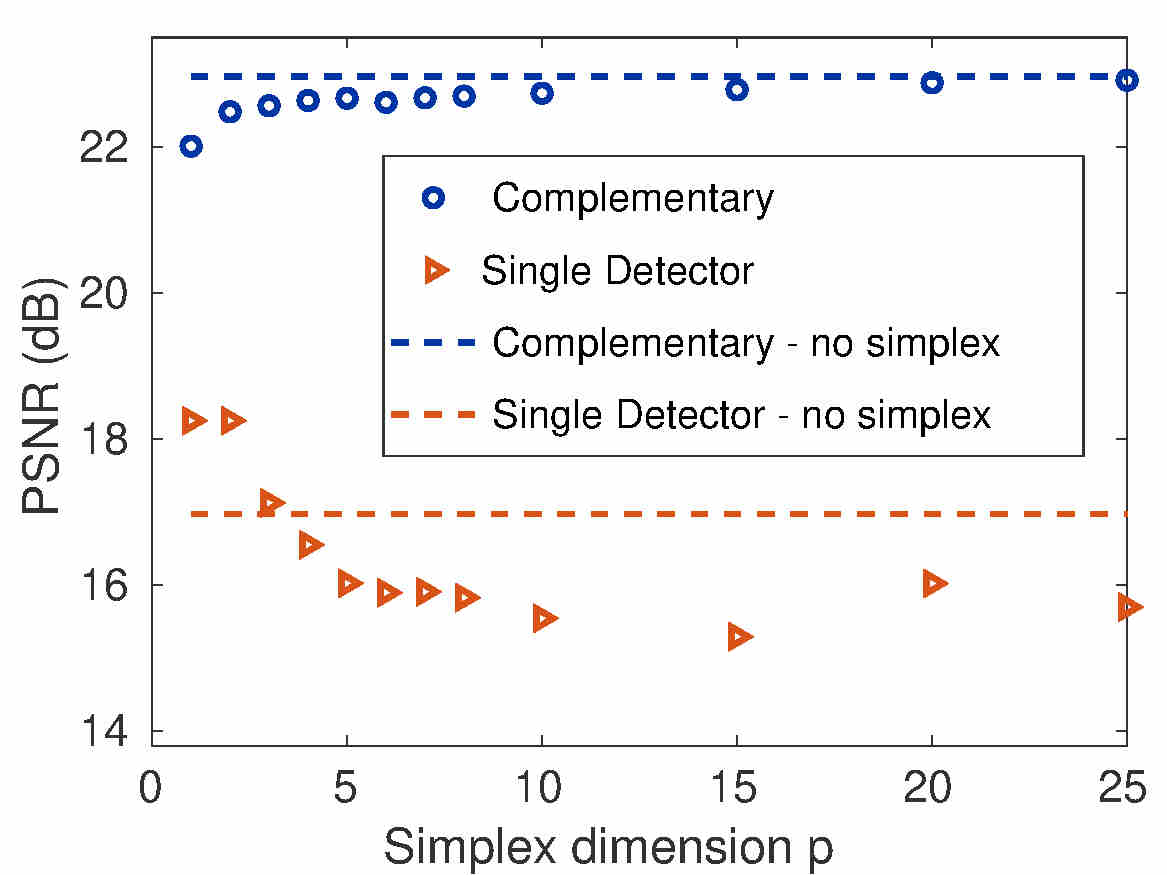}
\caption{Experimental PSNR with DCT sampling functions distributed over p-simplex with different values of $p$. Results obtained for a stationary scene with a complementary single-pixel camera (see Fig.~\ref{fig:experiment_scheme}), operating in 2 modes: either using one detector to measure the signal or both detectors for a differential measurement. 
For all values of $p$ the total number of displayed sampling functions equals $2000$, while the resolution of the patterns is $256\times 256$. 
We use dithering for binarization of the patterns before displaying them on DMD.
For comparison, the PSNRs with 2000 DCT functions without simplex method are also included (dashed lines). 
\label{fig:experiment_psnr}}
\end{figure}

\begin{figure}
\centering
\includegraphics[width=1\linewidth]{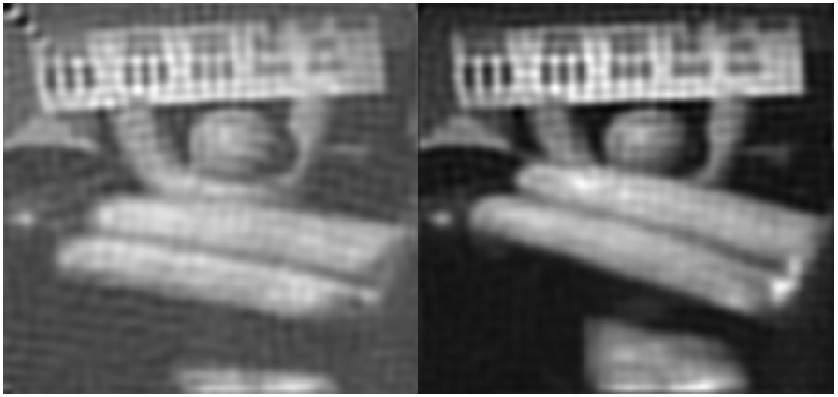}
\caption{Two exemplary video frames with pixel resolution $256\times 256$ captured in the same conditions by a single-pixel camera (complementary measurement mode) using DCT sampling patterns either directly (left) or distributed over a $10$-simplex (right). The number of sampling functions in both cases equals $2000$, which allows for $11$~fps video frame rate. The loss of contrast and artifacts present in the reconstruction obtained without the simplex method result from the varying bias over time in the measurement of a dynamic scene with changing brightness.
This effect is compensated by the use of the simplex method (see \href{https://www.osapublishing.org/ol/viewmedia.cfm?uri=ol-44-5-1241&seq=v001}{Visualization~1} and \href{https://www.osapublishing.org/ol/viewmedia.cfm?uri=ol-44-5-1241&seq=v002}{Visualization~2}). 
\label{fig:experiment_video}}
\end{figure}

Experimental PSNR values shown in Fig.~\ref{fig:experiment_psnr} as a function of $p$ have been obtained in an optical measurement with $K=2000$ binarized DCT sampling patterns with and without simplex coding. 
When only a single detector is used and the level of noise is large, simplex coding improves the PSNR, and gives the best results when $p\leq2$. When noise is reduced by complementary detection, simplex coding deteriorates the PSNR, due to the reduced compression ratio $k/n=K/(n+n\cdot p^{-1})$.
Still, when $p\geq10$ the results  obtained with and without simplex coding are similar. 
However, Fig.~\ref{fig:experiment_psnr} refers to the measurement in stationary conditions. When there are any moving objects in the scene, which affect the illumination balance between the two detectors in a complementary measurement, and the assumption of zero-mean noise can no longer be granted, the simplex coded sampling outperforms the direct approach. This is shown in Fig.~\ref{fig:experiment_video} and the respective \href{https://www.osapublishing.org/ol/viewmedia.cfm?uri=ol-44-5-1241&seq=v001}{Visualization~1} and \href{https://www.osapublishing.org/ol/viewmedia.cfm?uri=ol-44-5-1241&seq=v002}{Visualization~2}.

In this letter we propose to modify the measurement matrix for single-pixel imaging.
The same technique could be extended to other measurement methods based on structured illumination or spatial or temporal signal modulation with non-negative functions. The measurement becomes theoretically independent of the ambient illumination, and in practice becomes more robust to the varying conditions of the experiment. This method may be easily combined with complementary sampling and with linear image reconstruction methods such as FDRI. For a properly selected dimension of the simplex, the method improves the PSNR.

\textbf{Funding:} Narodowe Centrum Nauki (NCN) (UMO-2017/27/B/ST7/00885).




\bigskip

\bibliographystyle{unsrt}
\bibliography{report}


\end{document}